\documentclass[12pt]{article}

\usepackage{scicite}

\usepackage{graphicx}
\usepackage{times}
\usepackage{amsmath}

\newenvironment{sciabstract}{%
\begin{quote} \bf}
{\end{quote}}

\newcounter{lastnote}

\usepackage{setspace}

\title{Cascaded-mode interferometers: spectral shape and linewidth engineering}

\author{Jinsheng Lu$^1$, Ileana-Cristina Benea-Chelmus$^{1,2}$, Vincent Ginis$^{1,3}$, \\ Marcus Ossiander$^{1,4}$, Federico Capasso$^{1*}$ \\\\
\normalsize{$^{1}$Harvard John A. Paulson School of Engineering and Applied Sciences, 9 Oxford Street}\\
\normalsize{Cambridge, Massachusetts 02138, United States}\\
\normalsize{$^{2}$ Hybrid Photonics Laboratory, École Polytechnique Fédérale de Lausanne (EPFL)}\\
\normalsize{CH-1015, Switzerland}\\
\normalsize{$^{3}$Data Lab / Applied Physics, Vrije Universiteit Brussel 1050 Brussel, Belgium}\\
\normalsize{$^{4}$Institute of Experimental Physics, Graz University of Technology, 8010 Graz, Austria} 
\\\\
\normalsize{$^\ast$To whom correspondence should be addressed; E-mail: capasso@seas.harvard.edu.}\\
}

\date{}

\begin{document} 




\maketitle 

\doublespacing


\begin{sciabstract}
  Interferometers are essential tools to measure and shape optical fields, and are widely used in optical metrology, sensing, laser physics, and quantum mechanics. They superimpose waves with a mutual phase delay, resulting in a change in light intensity. A frequency-dependent phase delay then allows to shape the spectrum of light, which is essential for filtering, routing, wave shaping, or multiplexing. Simple Mach-Zehnder interferometers superimpose spatial waves and typically generate an output intensity that depends sinusoidally on frequency, limiting the capabilities for spectral engineering. Here, we present a novel framework that uses the interference of multiple transverse modes in a single multimode waveguide to achieve arbitrary spectral shapes in a compact geometry. Through the design of corrugated gratings, these modes couple to each other, allowing the exchange of energy similar to a beam splitter, facilitating easy handling of multiple modes. We theoretically and experimentally demonstrate narrow-linewidth spectra with independently tunable free spectral range and linewidth, as well as independent spectral shapes for various transverse modes. Our methodology can be applied to orthogonal optical modes of different orders, polarization, and angular momentum, and holds promise for sensing, optical metrology, calibration, and computing.
  \end{sciabstract}

\section*{Introduction}
The manipulation and control of the amplitude and phase of broadband light at each wavelength, known as optical spectral shaping, is fundamental for applications such as pulse shaping \cite{heritage1985picosecond,weiner1988high,weiner1995femtosecond,chou2003adaptive,cundiff2010optical,ye2011photonic,cao2021thermo}, microwave waveform generation through wavelength-to-time mapping of optical signals \cite{schnebelin2019programmable,khan2010ultrabroad,yao2010arbitrary,zhang2020photonic,ghelfi2014fully,marpaung2013integrated,fandino2017monolithic,marpaung2019integrated,liu2020integrated}, and sensing in biochemistry, medicine, and physics \cite{fan2008sensitive,albert2013tilted,tan2013microfiber,klimov2015chip,monzon2011compact,li2021chip,schreiber2023variations}. The first attempt to manipulate the optical spectrum dates back to Newton’s prism experiments \cite{newton1993new}, where white light was decomposed into its constituent colors. Building on Newton’s work, researchers have then implemented a spatial mask or spatial light modulator to control the amplitude (and possibly the phase) of each of these colors \cite{heritage1985picosecond,weiner1988high,weiner1995femtosecond,chou2003adaptive,cundiff2010optical}. This parallel manipulation offers spectral control with a frequency resolution limited by the pixel size of the mask and the beam diameter at the mask, and requires large components and space, making miniaturization challenging. 

\begin{figure}[h!]
\includegraphics[width = 0.75\textwidth]{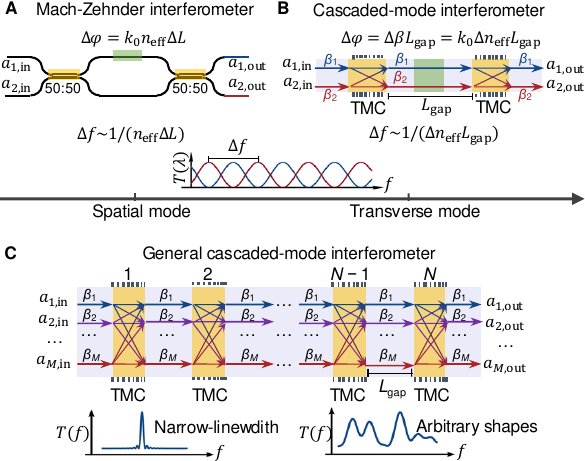}
\centering
\caption{\label{fig1}{\bf Concept of cascaded-mode interferometers.} ({\bf A}) A typical Mach–Zehnder interferometer (MZI) with two inputs and two outputs. The relative phase shift between the two arms is ${\Delta \varphi}$ = ${k_0 n_\mathrm{eff} \Delta L}$, where ${k_0}$, ${n_\mathrm{eff}}$, and ${\Delta L}$ are the wavenumber, the effective index of the spatial mode, and the length difference of the two arms, respectively. The output interference spectrum is shown below and its free spectral range (${\Delta f}$) is proportional to ${1/(n_\mathrm{eff} \Delta L)}$. Note that the group index $n_{g}$ should be used here if dispersion is considered. ({\bf B}) A cascaded-mode interferometer, as a counterpart to the MZI. It consists of two orthogonal modes (propagation constants ${\beta_1}$, ${\beta_2}$) in a multimode waveguide and two transmissive mode{convert}ers (TMCs) separated by a distance ${L_\mathrm{gap}}$. The relative phase shift between the two modes is ${\Delta \varphi}$ = ${k_0 \Delta n_\mathrm{eff} L_\mathrm{gap}}$, where ${\Delta n_\mathrm{eff}}$ is the effective index difference of the modes. Its free spectral range of the output spectrum is proportional to ${1/(\Delta n_\mathrm{eff} L_\mathrm{gap})}$. ({\bf C}) A general cascaded-mode interferometer, where multiple orthogonal modes with propagation constants ${\beta_j}$ are converted and mixed by multiple TMCs separated by ${L_\mathrm{gap}}$. ${a_{j,\mathrm{in}}}$ and ${a_{j,\mathrm{out}}}$ (${j = 1, 2, ..., M}$) represent the amplitudes of input and output mode ${j}$, respectively. A narrow linewidth spectrum and a spectrum with arbitrary shapes generated by suitably designed cascaded-mode interferometers are shown below. }
\end{figure}

Simple filtering functions can be implemented via a simple Mach-Zehnder interferometer (Fig. \ref{fig1}A) \cite{choi2008miniature}, that splits and recombines two beams (denoted as ${a_{1,\mathrm{in}}}$ and ${a_{2,\mathrm{in}}}$) after sending them along two paths that differ by a length ${\Delta L}$. Such interferometers are routinely implemented in integrated photonics, utilizing on the platform's ability to realize compact splitters and waveguides of arbitrary length. However, the spectral response depends sinusoidally on frequency. Consequently, the output power spectrum of the two output beams (denoted as ${a_{1,\mathrm{out}}}$ and ${a_{2,\mathrm{out}}}$) oscillates with a periodicity ${\Delta f \propto 1/\left({n_\mathrm{eff}\Delta L}\right)}$ that depends on the effective index ${n_\mathrm{eff}}$ of the waveguide mode and the path length difference ${\Delta L}$ (we do not consider dispersion here for simplicity). Bragg gratings provide finer control that achieves wavelength-specific and bandwidth-controlled reflection or filtering through interference of an infinity of waves \cite{zhang2018fully,kim2003optical}. Such narrow-linewidth response relies strictly on invoking more than two waves and can also be achieved using multilayer thin-films \cite{macleod2010thin,yang2022wavelength,wan2021switchable}, Fabry-Perot interferometers \cite{choi2008miniature}, arrayed waveguide gratings \cite{Cheben:07}, and fiber interferometers \cite{gu1998wavelength,geng2013line,mohammed2006all,Antonio-Lopez:10}. More complex spectral manipulation can be achieved using on-chip spectral shapers, typically consisting of multiple resonators, which offer high spectral resolution and programmability \cite{schnebelin2019programmable,khan2010ultrabroad,yao2010arbitrary,zhang2020photonic,wu2018advanced,cohen2018response}. 

The working principle of most optical devices mentioned above relies on the interference of beams that are reflected multiple times. However, their amplitudes are constrained by the reflection or transmission coefficients of the mirrors or interfaces, and the phases are limited by the propagation lengths and propagation constants, which are integer multiples of the cavity length or thin-film thickness. This typically results in the amplitudes being dependent on each other, leaving the requirement for independent control unaddressed. Furthermore, the propagation constants of these beams are typically the same. 

In this work, we propose to shape the spectra of light by using an alternative to conventional Mach-Zehnder interferometers: we exploit multiple transverse modes of a multimode waveguide on silicon on insulator (SOI) platform, instead of the spatial modes of two individual waveguides (Fig. \ref{fig1}A). To couple these modes, we use transmissive mode converters (TMC) that transfer energy from one mode to another, depending on the so-called splitting ratio, similar to a beam splitter. This approach enables a similar spectral shaping when using two transmissive mode converters in a geometry as shown in Fig. \ref{fig1}B, albeit with a periodicity that is not determined by a path imbalance but instead by an imbalance in the propagation constants (${\beta_i = \frac{2\pi}{\lambda} n_{\mathrm{eff},i}}$, ${i}$ = 1, 2) of the two modes ${\Delta f \propto 1/\left(\Delta n_\mathrm{eff} L_\mathrm{gap}\right)}$, with ${\Delta n_\mathrm{eff}}$ being the difference in the effective refractive index of the employed transverse modes and ${L_\mathrm{gap}}$ being the length of the multimode waveguide between the mode converters. We show that this compact implementation provides a straightforward extension to cascading more mode converters ($N$) and a higher number of modes ($M$) with propagation constants ${\beta_1}$, ${\beta_2}$, ..., ${\beta_M}$ (Fig. \ref{fig1}C). Building upon this concept, we develop and present a generalized framework that computes the exact spectra of multiple interfering transverse modes through transfer matrix formalism, and their dependency on the splitting ratio of the mode converters. We demonstrate narrow-linewidth (i.e., high finesse) and arbitrary spectra using our cascaded-mode interferometers. One of the promising aspects of our device is that, unlike traditional technologies such as Fabry-Perot interferometers, its finesse remains unaffected by losses, enabling the integration of switchable or active materials without compromising the device's spectral performance.

\section*{Spectral engineering using cascaded-mode interferometers}

In the case of $N$ transmissive mode converters spaced by the same distance ${L_\mathrm{gap}}$ depicted in Fig. \ref{fig1}C, the amplitudes of the modes at the output of the interferometer depend on the input modes and the transfer function of the various components as
\begin{equation}\label{eq1}
   {\mathbf{a_\mathrm{out}(\lambda)} = \mathbf{T_c}(\mathbf{T_{wg}}\mathbf{T_c})^{N-1}\mathbf{a_\mathrm{in}}}
\end{equation}
where ${\mathbf{a_\mathrm{in}} = \left(a_{1,\mathrm{in}},a_{2,\mathrm{in}},...,a_{M,\mathrm{in}}\right)^T}$ and ${\mathbf{a_\mathrm{out}} = \left(a_{1,\mathrm{out}},a_{2,\mathrm{out}},...,a_{M,\mathrm{in}}\right)^T}$ are the amplitude arrays of the input and output modes of the interferometer. This is only true for pure forward-scattering converters. ${\mathbf{T_c}}$ and ${\mathbf{T_{wg}}}$ are the transmittance matrix of the mode converter and the multimode waveguide between the mode converters, respectively. The mode ${j}$ accumulates a phase term ${H_j(\lambda)}$ = ${e^{-i \beta_j L_\mathrm{gap}}}$ = ${e^{-i \frac{2\pi}{\lambda}n_{\mathrm{eff},j} L_\mathrm{gap}}}$ (${j = 1, 2, ..., M}$) during propagation in the multimode waveguide between the mode converters. Therefore, ${\mathbf{T_{wg}}}$ represents a propagation phase matrix, which can be written as
\begin{equation}\label{eq2}
    \mathbf{T_{wg}} = 
    \begin{pmatrix}
        {H_1(\lambda)}  & 0  & ...  & 0 \\
        0 & {H_2(\lambda)} & ... & 0 \\
        ... & ... & ... & ... \\
        0 & 0 & ... & {H_M(\lambda)}
    \end{pmatrix}
\end{equation}

The transmittance matrix ${\mathbf{T_c}}$ representing an arbitrary mode conversion (or so-called beam-splitting) function is given by

\begin{equation}\label{eq3}
    \mathbf{T_c} = 
    \begin{pmatrix}
       { t_{11}}  &  {t_{12}}  & ...  &  {t_{1M}} \\
         {t_{21}} &  {t_{22}} & ... &  {t_{2M}} \\
        ... & ... & ... & ... \\
         {t_{M1}} &  {t_{M2}} & ... & {t_{MM}}
    \end{pmatrix}
\end{equation}
where ${t_{ij}}$ (${i,j = 1, 2, ..., M}$) is the transmittance coefficient indicating the mode conversion from mode ${j}$ to mode ${i}$ in a transmissive way. $\mathbf{T_c}$ and $\mathbf{T_{wg}}$ are unitary matrices if there is no loss. $\mathbf{T_c}$ is a symmetric matrix if the mode conversion is reciprocal. 

We consider a case that ${N}$ = 2 to show the ability of arbitrary spectral shaping of the cascaded-mode interferometer. Equation (\ref{eq1}) can then be simplified to $\mathbf{a_\mathrm{out}}$ = ${\mathbf{T_c}\mathbf{T_{wg}}\mathbf{T_c}\mathbf{a_\mathrm{in}}}$. Combined with Eqs. (\ref{eq2}) and (\ref{eq3}), the amplitude spectrum of mode ${j}$ in the output of the cascaded-mode interferometer can be calculated as
\begin{equation}\label{eq4}
    {a_{j,\mathrm{out}}(\lambda) = \sum_{m=1}^{M}\sum_{n=1}^M  t_{jn} e^{-i \frac{2\pi}{\lambda}n_{\mathrm{eff},n} L_\mathrm{gap}} t_{nm} a_{m,\mathrm{in}}}
\end{equation}

It is important to note that the effective index ${n_{\mathrm{eff},n}}$ in this series is not freely selectable but is instead restricted to specific values, typically non-equidistant, determined by the waveguide cross-section. Despite these non-equidistant ${n_{\mathrm{eff},n}}$ values, the series can still effectively approximate a wide range of predetermined functions, similar to a standard Fourier series \cite{Vincent_Science21}. Therefore, by designing the transmittance coefficients ${t_{ij}}$, we can achieve nearly arbitrary spectral shapes of the output modes.

\begin{figure}[h!]
\includegraphics[width = \textwidth]{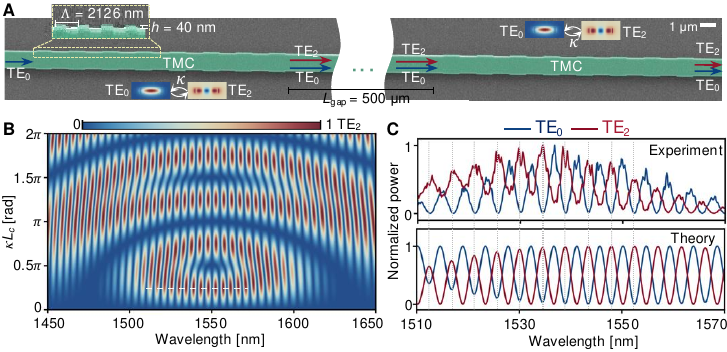}
\centering
\caption{\label{fig2}{\bf Interference spectra generated by a cascaded-mode interferometer.} ({\bf A}) Scanning electron microscope image of a fabricated cascaded-mode interferometer comprising two waveguide transverse modes ($\mathrm{TE_0}$ and $\mathrm{TE_2}$) and two transmissive mode converters (TMCs) separated by a distance ${L_\mathrm{gap}}$ = 500 ${\mu}$m. The TMCs are made of nanogratings with period ${\Lambda}$ = 2126 nm, the length ${L_c}$ = 8${\Lambda}$ = 17 ${\mu}$m, and the corrugated grating depth ${h}$ = 40 nm. The width of the multimode waveguide is 1100 nm. The aspect ratio of the zoomed-in image in (A) is set to 1:6 to visualize the nano-gratings better. ({\bf B}) Calculated output power spectra of the $\mathrm{TE_2}$ modes for varying coupling strength ${\kappa L_c}$ which is the product of the nanograting's coupling coefficient (${\kappa}$) and the grating length (${L_c}$). The white dashed lines in (B) correspond to the red curve (theory) in (C) where ${\kappa L_c}$ = 0.25${\pi}$, that is, a mode power splitting ratio of 50:50. ({\bf C}) Measured and calculated output power spectra of the $\mathrm{TE_0}$ and $\mathrm{TE_2}$ modes when ${L_\mathrm{gap}} = 500$ ${\mu}$m and ${\kappa L_c}$ = 0.25${\pi}$. The input mode is $\mathrm{TE_0}$ mode. }
\end{figure}

\section*{On-chip cascaded-mode interferometers}

In the first experiment, we realize the most simple on-chip cascaded-mode interferometer, featuring two input modes and two output modes which we choose to be $\mathrm{TE_0}$ and $\mathrm{TE_2}$ as shown in Fig. \ref{fig2}A. We corrugate the multimode waveguide with a periodicity of ${\Lambda}$. This nanograting provides a momentum at the central wavelength ${\lambda_0}$ to satisfy the phase matching (${\Lambda}$ = ${{\lambda_0}/{(n_{\mathrm{eff},1}- n_{\mathrm{eff},2})}}$ = ${{\lambda_0}/\Delta n_\mathrm{eff}}$) for co-directional coupling between the $\mathrm{TE_0}$ and $\mathrm{TE_2}$ mode. The transmittance coefficients ${t_{ij}}$ (${i,j}$ = 1, 2) in the transmittance matrix ${\mathbf{T_c}}$ of this mode converter can be derived from coupled mode theory \cite{huang1994coupled} (section S1) as ${t_{11}},{t_{22}}$ = ${ \cos{(s L_c)}\mp i\frac{\delta}{s}\sin{(s L_c)}}$ and ${t_{12}}$ = ${t_{21}}$ = ${-i\frac{\kappa}{s}\sin{(s L_c)}}$, where ${\delta}$ = ${\frac{\pi \Delta n_\mathrm{eff}}{\lambda} - \frac{\pi}{\Lambda}}$ is the phase mismatch, ${\kappa}$ is the coupling coefficient, ${L_c}$ is the grating length, and ${s = \sqrt{\delta^2+\kappa^2}}$. Phase matching is only achieved at the central wavelength, that is, ${\delta(\lambda = \lambda_0)} = 0$, and the phase mismatch is proportional to the deviation of the wavelength from the central wavelength: ${\delta(\lambda) \approx \frac{\pi\Delta n_\mathrm{eff}(\lambda-\lambda_0)}{\lambda_0^2}}$. The power conversion efficiency of the mode converter is ${\eta = \frac{\kappa^2}{\kappa^2+\sigma^2}\sin^2(sL_c)}$. The bandwidth of the mode converter (${\Delta\lambda_\mathrm{BW}}$), determined by this term ${\frac{\kappa^2}{\kappa^2+\sigma^2}}$, can be derived as ${\Delta\lambda_\mathrm{BW} = \frac{2\kappa \lambda_0^2}{\pi\Delta n_\mathrm{eff}}}$, which increases with ${\kappa}$. Suppose we input $\mathrm{TE_0}$ mode into this cascaded-mode interferometer, that is, ${\mathbf{a_\mathrm{in}}}$ = ${[1,0]^T}$, the amplitude of the $\mathrm{TE_0}$ mode at the output can be calculated using Eq. (\ref{eq4}) as
\begin{equation}\label{eq5}
    {a_{1,\mathrm{out}} = \left(\cos(s L_c)-i\frac{\delta}{s}\sin(s L_c)\right)^2 e^{-i\frac{2\pi}{\lambda} n_{\mathrm{eff},1} L_\mathrm{gap}} - \frac{\kappa^2}{s^2}\sin^2(s L_c) e^{-i\frac{2\pi}{\lambda} n_{\mathrm{eff},2} L_\mathrm{gap}} }
\end{equation}
We have ${|a_{1,\mathrm{out}}|^2+|a_{2,\mathrm{out}}|^2}$ = 1 according to the conservation of energy. The above equation can be further simplified to be ${a_{1,\mathrm{out}} = \cos^2(\kappa L_c) e^{-i\frac{2\pi}{\lambda} n_{\mathrm{eff},1} L_\mathrm{gap}} - \sin^2(\kappa L_c) e^{-i\frac{2\pi}{\lambda} n_{\mathrm{eff},2} L_\mathrm{gap}} }$, supposing that the phase match is satisfied for all wavelengths (${\delta} = 0$). To achieve a sinusoidal modulation of the power transmitted through the interferometer as a function of frequency with maximal visibility, the two transmissive mode converters need to split the power equally into $\mathrm{TE_0}$ and $\mathrm{TE_2}$, in analogy to 50:50 beam splitters used in conventional Mach-Zehnder modulators (fig. S7). Therefore, the coupling strength ${\kappa L_c}$ should be equal to ${\frac{\pi}{4}}$, leading to ${a_{1,\mathrm{out}} = \frac{1}{2} (e^{-i\frac{2\pi}{\lambda} n_{\mathrm{eff},1} L_\mathrm{gap}} - e^{-i\frac{2\pi}{\lambda} n_{\mathrm{eff},2} L_\mathrm{gap}}) }$ and a maximal visibility of the interference spectrum. 

\begin{figure}[h!]
\includegraphics[width = \textwidth]{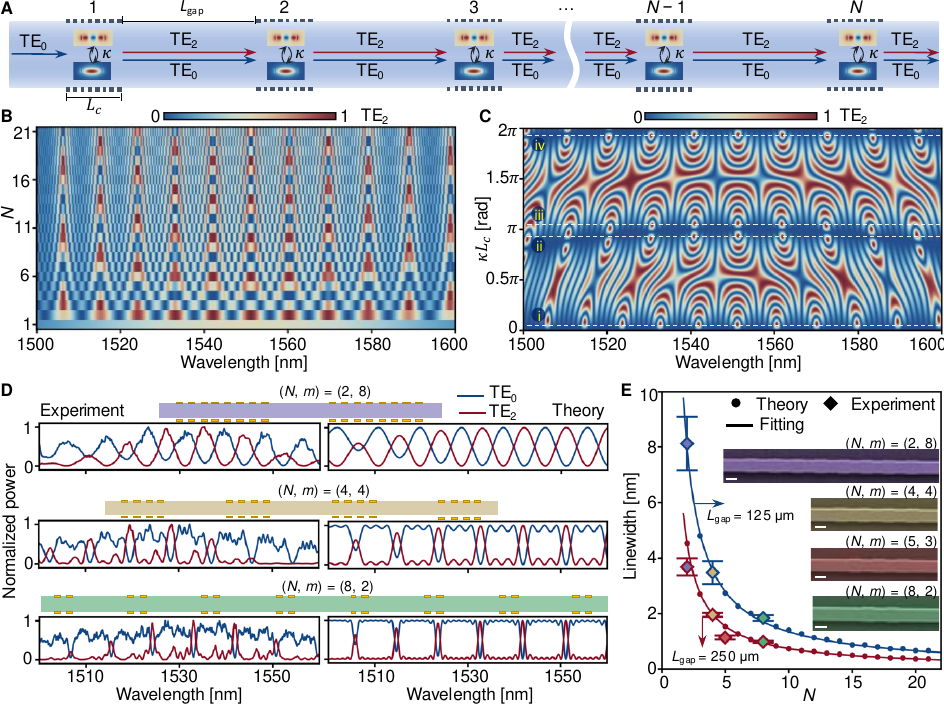}
\centering
\caption{\label{fig3}{\bf Cascaded-mode interferometer with multiple TMCs for precise tuning of FSR and linewidth.} ({\bf A}) Schematic of a cascaded-mode interferometer with multiple TMCs and two modes ($\mathrm{TE_0}$ and $\mathrm{TE_2}$). ({\bf B}) Calculated output power spectra of the $\mathrm{TE_2}$ mode with varying the number (${N}$) of TMCs at ${\kappa L_c = 0.25\pi}$. Parameters used in the numerical calculations: ${L_\mathrm{gap}}$ = 250 ${\mu}$m, ${a_{1,\mathrm{in}}}$ = 1, ${a_{2,\mathrm{in}}}$ = 0, ${n_{1,\mathrm{eff}}}$ = 2.74, ${n_{2,\mathrm{eff}}}$ = 1.98. ({\bf C}) Calculated output power spectra of the $\mathrm{TE_2}$ mode for varying coupling strength ${\kappa L_c}$ when ${N}$ = 8. The four dashed lines in (C) represent the positions where (i) ${\kappa L_c = 0.5\pi/N}$, (ii) ${\kappa L_c = (N-0.5)\pi/N}$, (iii) ${\kappa L_c = (N+0.5)\pi/N}$, and (iv) ${\kappa L_c = (2N-0.5)\pi/N}$ with ${N = 8}$. ({\bf D}) Measured (left) and calculated (right) output power spectra of the two modes for different $N$ with coupling strengths ${\kappa L_c = 0.5\pi/N}$ (${N = 2, 4, 8}$). The multiplication of the number of the mode converters ${N}$ and the number of the grating period of a single mode converter ${m}$ equals 16. ({\bf E}) Measured and calculated linewidth variation with ${N}$. Fitting equations: ${12.88/(N-0.38)}$ and ${7.02/(N-0.44)}$ for ${L_\mathrm{gap}}$ = 125 $\mu$m and ${L_\mathrm{gap}}$ = 250 ${\mu}$m, respectively. The input mode is $\mathrm{TE}_0$ mode. The inserts are the scanning electron microscope images of the mode converters of the fabricated cascaded-mode interferometer device at different ${N}$. The scale bars are 1 ${\mu}$m.}
\end{figure}

\begin{figure}[h!]
\includegraphics[width = 0.9\textwidth]{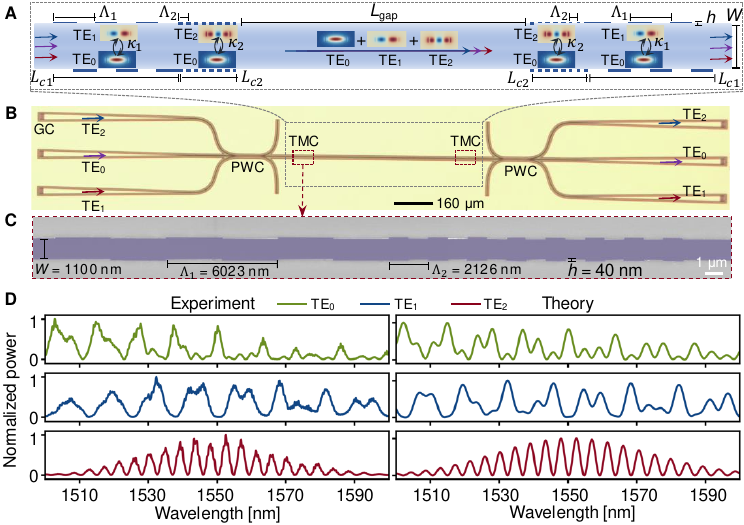}
\centering
\caption{\label{fig4}{\bf Cascaded-mode interferometer with multiple waveguide modes for parallel spectra engineering.} ({\bf A}) Schematic of a cascaded-mode interferometer with three waveguide modes ($\mathrm{TE_0}$, $\mathrm{TE_1}$, and $\mathrm{TE_2}$) and two TMCs separated by a distance of ${L_\mathrm{gap}}$. The TMCs consist of two sets of nanogratings. One set of the nanograting is asymmetric and with the period ${\Lambda_1}$ = 6023 nm, the length ${L_{c1}}$ = 3${\Lambda}$ = 18 ${\mu}$m, and the coupling strength ${\kappa_1 L_{c1}}$ = 0.19${\pi}$, used for mode conversion between $\mathrm{TE_0}$ and $\mathrm{TE_1}$. The other set of grating is symmetric and with the period ${\Lambda_2}$ = 2126 nm, the length ${L_{c2}}$ = 8${\Lambda}$ = 17 ${\mu}$m, and the coupling strength ${\kappa_2 L_{c2}}$ = 0.25${\pi}$, used for mode conversion between $\mathrm{TE_0}$ and $\mathrm{TE_2}$. The multimode waveguide width is ${W}$ = 1100 nm. ${L_\mathrm{gap}}$ = 500 ${\mu}$m. The grating depth is 40 nm. ({\bf B}) Optical image of the fabricated device. GC: grating coupler, PWC: parallel waveguide coupler, TMC: transmissive mode converter. ({\bf C}) Scanning electron microscope image of the left TMC in (B). ({\bf D}) Measured and calculated output power spectra of $\mathrm{TE_0}$, $\mathrm{TE_1}$, and $\mathrm{TE_2}$ modes when the input mode is $\mathrm{TE_0}$ mode.}
\end{figure}

Figure. \ref{fig2}B depicts the calculated wavelength-dependent transmitted power contained in mode $\mathrm{TE_2}$ after the interferometer upon sending $\mathrm{TE_0}$ into the interferometer, as a function of coupling strength ${\kappa L_c}$ for a gap length ${L_\mathrm{gap} = 500~\mu}$m. We notice that the lowest coupling strength for maximal visibility is when ${\kappa L_c = \frac{\pi}{4}}$ (white dashed cut line). Operating the cascaded-mode interferometer at this point allows to use the shortest length for the mode converter, which is beneficial for the footprint of the device, or, alternatively, the smallest corrugation. The bandwidth of the mode converter is limited by phase mismatch, which becomes detrimental as soon as $\delta$ becomes a significant portion of $s$. As follows from Eq. (\ref{eq5}), the bandwidth can be increased by increasing the coupling strength to, for example, ${\kappa L_c = \frac{3\pi}{4}}$ or ${\kappa L_c = \frac{5\pi}{4}}$. In these cases, the larger coupling strength compensates for the phase mismatch, although at the expense of longer gratings or larger corrugations. To validate our concept,  we experimentally report the wavelength-dependent power of the output modes $\mathrm{TE_0}$ and $\mathrm{TE_2}$ after the interferometer, under an input $\mathrm{TE_0}$ mode (upper graph of Fig. \ref{fig2}C). We use a grating period $\Lambda$ = 2126 nm. The coupling coefficient of the mode conversion between the $\mathrm{TE_0}$ and $\mathrm{TE_2}$ modes is ${\kappa}$ = $\frac{\pi}{32\Lambda}$ = 0.046 ${\mu \mathrm{m}^{-1}}$. We observe alternating powers that match well with our analytical model (lower graph of Fig. \ref{fig2}C and fig. S7) as well as simulation results (fig. S8), in line with expectations from a Mach-Zehnder interferometer. Note that the envelope of the measured spectra in Fig. \ref{fig2}C (also in Fig. \ref{fig3}D and Fig. \ref{fig4}D) results from the parallel waveguide coupler (section S2 and fig. S5), which is used to load and unload the modes from the multimode waveguide. The free spectral range of the interference spectra is measured as 19.4 nm, 8.5 nm, and 4.4 nm when ${L_\mathrm{gap}}$ = 100 ${\mu}$m, 250 ${\mu}$m, and 500 ${\mu}$m, respectively, which are in good agreement with the calculations: 21.5 nm, 8.6 nm, and 4.3 nm using the formula ${\Delta \lambda_\mathrm{FSR} = \frac{\lambda_0^2}{\left(n_{g,1}-n_{g,2}\right)L_\mathrm{gap}}}$, where ${n_{g,1}}$= 4.85 and ${n_{g,2}}$ = 3.75 are the group indexes of the $\mathrm{TE_0}$ and $\mathrm{TE_2}$ modes at the central wavelength ${\lambda_0}$ = 1538 nm (fig. S9). 

In many optical applications, controlling light's spectrum to achieve a narrower transmission linewidth is desirable, for example, in filtering or routing. In the following, we will show that cascading several mode converters provide a useful knob to achieve this, as visualized in Fig. \ref{fig3}A. A tempting approach could be to simply cascade $N$ 50:50 mode converters along a single multimode waveguide. However, we find from computing the total transfer function of the system using Eq. (\ref{eq1}) that concatenating $N$ 50:50 mode converters leads to a narrowing of the transmission spectra, at the expense of multiple undesired sidebands (Fig. \ref{fig3}B and fig. S10). This is typically not desired in filtering and routing applications which rely on achieving a vanishingly small insertion loss only at one desired wavelength, and close to zero transmission elsewhere. Guided by our mathematical derivation (section S3), we find that, instead, the optimal coupling strength of the grating needs to generally satisfy the condition ${\kappa L_c = \frac{\pi}{2N}}$, with $N$ the total number of mode converters. In this case, the corresponding splitting ratio (SR) of the mode converter at the central wavelength is ${SR = \frac{\eta}{1-\eta}}$, where ${\eta = \sin^2(\frac{\pi}{2N})}$. For $N$ = 8, the mode converter has a splitting ratio of around 0.04, which means that when 1 mW $\mathrm{TE_0}$ mode is input into the mode converter, it gets 0.038 mW $\mathrm{TE_2}$ mode and 0.962 mW $\mathrm{TE_0}$ mode. We exemplify this by reporting in Fig. \ref{fig3}C the total transmitted power for $\mathrm{TE_2}$ and ${N = 8}$ for various coupling strengths ${\kappa L_c}$. As expected, at ${\kappa L_c = \frac{\pi}{2\times 8}}$, the transmitted power features narrow-linewidth transmission (dashed line (i)). Other coupling strengths where this is satisfied are ${\kappa L_c = \pi \pm \frac{\pi}{2N}}$, ${\kappa L_c = 2\pi \pm \frac{\pi}{2N}}$ and so on (${N}$ = 8). 

We fabricated a series of cascaded-mode interferometers with ${N}$ = 2, 4, and 8 while adjusting the number of periods to satisfy ${\kappa L_c = \frac{\pi}{2N}}$ for each one of them. We can easily get ${m \times N = \frac{\pi}{2\kappa \Lambda}}$ = 16 according to ${L_c = m\Lambda}$ and ${\kappa = \frac{\pi}{32\Lambda}}$. Consequently, the number of the grating periods of the mode converter is ${m}$ = 8, 4, and 2, respectively. The maximum achievable $N$ is 8 in this case (because ${m \times N}$ = 16 and $m \geq 2$). However, by utilizing a multimode waveguide with a larger width or a mode conversion grating with weaker corrugation, we can reduce the coupling coefficient (${\kappa}$), thereby enabling a significantly larger $N$. For example, the maximum achievable $N$ becomes 34 when the width of the multimode waveguide and the corrugation depth of the grating are 2500 nm and 20 nm, respectively (fig. S23). Note that in the condition of ${\kappa L_c < \frac{\pi}{2N}}$, we can still obtain narrow linewidth spectra but with a smaller amplitude at the power spectrum peaks, which is determined by $|a_\text{peak}|^2 = \sin^2(\kappa L_c N)$ (fig. S23). 

In line with our modeling, we experimentally find that an increased number of converters significantly reduce the linewidth of the transmission spectra while efficiently suppressing its off-resonance transmission, as shown in Fig. \ref{fig3}D. We find the experimental linewidth to match well with theory, which is corroborated by two sets of devices with different gap lengths ${L_\mathrm{gap} = 125~\mu}$m (fig. S15) and ${L_\mathrm{gap} = 250~\mu}$m (Fig. \ref{fig3}D). The free spectral range of the narrow linewidth spectra generated by this cascaded-mode interferometer, in this case, can be decreased by increasing the gap between TMCs (${\Delta \lambda_\mathrm{FSR} \propto \frac{1}{L_\mathrm{gap}}}$) (fig. S14), and the linewidth (full width at half maximum) ${\Delta \lambda_\mathrm{FWHM}}$ decreases when increasing the number ${N}$ of TMCs (Fig. \ref{fig3}, fig. S11-13), which can be approximated by ${\frac{\Delta \lambda_\mathrm{FSR}}{N}}$. A more accurate expression would be ${\Delta \lambda_\mathrm{FWHM}}$ = ${\frac{\Delta \lambda_\mathrm{FSR}+c_1}{N+c_2}}$, where ${c_1}$ and ${c_2}$ are fitting coefficients (Fig. \ref{fig3}E). The finesse of this cascaded-mode interferometer, defined as the ratio of the free spectral range and the linewidth, therefore, is approximately equal to the number of the mode converters: ${F = \frac{\Delta \lambda_\mathrm{FSR}}{\Delta \lambda_\mathrm{FWHM}} \approx N}$. We note here however that, unlike in the case of resonators, the finesse is not related to a field enhancement, but rather to the contrast of the transmission spectrum in a given band. In contrast to the finesse of a traditional Fabry-Perot interferometer that is sensitive to loss, the finesse of our cascaded-mode interferometer is loss-independent, whereas the total transmitted power is loss-dependent (fig. S22). 

\section*{Multi-dimensional on-chip cascaded-mode interferometers}

Shaping the spectrum of a light source simultaneously in multiple ways is a requirement in many applications. For example, having the ability to flatten part of the spectrum can be beneficial for spectroscopy over a broad bandwidth, and simultaneously modulating another part of it in intensity can be important for achieving more complex time-domain profiles. A large ${\frac{\Delta \lambda_\mathrm{FSR}}{\Delta \lambda_\mathrm{FWHM}}}$ is also often needed to cut fundamental radiation in spectroscopy or on-chip generated frequency combs. 

We show in the following that such manipulation of light's spectrum can be accomplished by extending the cascaded-mode interferometer of Fig. \ref{fig2} to provide efficient conversion between not only two but more distinct transverse modes. In this case, the spectral profile of two orthogonal modes can be arbitrarily shaped by dumping the remaining energy into the third mode (fig. S16-21). To exemplify this concept, we resort to mode converters that feature a pair of periodicities, as to convert $\mathrm{TE_0}$ into $\mathrm{TE_1}$ with a coupling strength $\kappa_1$ and $\mathrm{TE_0}$ into $\mathrm{TE_2}$ with a coupling strength $\kappa_2$, as seen in Fig. \ref{fig4}A. Figure. \ref{fig4}B depicts an optical microscope image of the fabricated device, showcasing straight-forward multiplexing and de-multiplexing of the various transverse modes using mode-selective parallel waveguide couplers (section S2, fig. S5, and fig. S6). Coupling $\mathrm{TE_0}$ with $\mathrm{TE_1}$ requires a grating that is asymmetrically displaced on the two sides of the waveguide due to the different symmetry of the two modes, whereas coupling $\mathrm{TE_0}$ with $\mathrm{TE_2}$ requires a symmetric one, as visible from the scanning electron microscope image in Fig. \ref{fig4}C. By choosing ${\kappa_1 L_{c1}= 0.19\pi}$ and ${\kappa_1 L_{c1}= 0.25\pi}$, we experimentally confirm in Fig. \ref{fig4}D (left panel) (also in fig. S18) that significantly different spectral shapes can be achieved for the three transverse modes, in line with our analytical description (Fig. \ref{fig4}D, right panel).

\section*{Discussion and outlook}
In summary, we demonstrate how interference of multiple spatial modes in a single interferometer can be utilized to control the spectral response of light using transmissive mode converters. Our approach leverages the capabilities of photonics to engineer the effective index and propagation constant through waveguide design, and the conversion bandwidth and efficiency through corrugation. By propagating multiple modes with different propagation constants within a single compact multimode waveguide, we circumvent the need for multiple waveguides that are otherwise used in multi-armed interferometers. By designing index-matched parallel waveguide couplers, we demultiplex these transverse modes. Altogether, this enables a smaller footprint and greater design flexibility than the traditional Mach–Zehnder interferometer. 

Our cascaded-mode interferometer shows excellent performance in spectral shaping. One interesting example we demonstrated is spectra with narrow transmission peaks or valleys whose linewidth and FSR are determined by the number of mode converters and the gap distance between them, respectively. The narrow-linewidth spectra can be applied in optical fiber sensing with advanced sensitivity compared with previous interferometric fiber optic sensors \cite{lee2012interferometric}. Our cascaded-mode interferometer works in a transmissive way without reflections. Therefore, the light energy is distributed in the entire device, which differs from the Fabry-Perot resonators where energy is built up in the cavity. In this regard, the Fabry-Perot resonators are more suitable for sensing in a local area, whereas the cascaded-mode interferometer device has its advantages in distributed sensing \cite{needham2024label} and is more robust to optical loss. Our approach is more flexible in spectral shaping compared to alternative solutions proposed using transmissive Bragg gratings on the silicon nitride platform \cite{mohanty2017quantum}, or multiarmed interferometers in fibers \cite{weihs1996all}. As such, they can become a tool for on-chip quantum interference between many modes \cite{crespi2013integrated,seron2023boson}. Our cascaded-mode interferometers can be programmable by using tunable mode converters with thermo- or electro-optical effects, demonstrating the potential for integrated, space-efficient optical computing applications \cite{zhou2022photonic,zhu2022space,meng2023compact,du2024ultracompact}.

Our studies offer a generalized theory framework for spectral shaping, opening up new directions for sensing, wavelength-isolation filtering, waveform shaping, and narrow-linewidth light amplifying, and inspiring various research on interference-based optical engineering and beyond optics.

\bibliography{scibib}

\begin{thebibliography}{10}

\bibitem{heritage1985picosecond}
J.~Heritage, A.~Weiner, R.~Thurston, {\it Opt. Lett.\/} {\bf 10}, 609 (1985).

\bibitem{weiner1988high}
A.~M. Weiner, J.~P. Heritage, E.~Kirschner, {\it J. Opt. Soc. Am. B\/} {\bf 5},
  1563 (1988).

\bibitem{weiner1995femtosecond}
A.~M. Weiner, {\it Prog. Quantum Electron.\/} {\bf 19}, 161 (1995).

\bibitem{chou2003adaptive}
J.~Chou, Y.~Han, B.~Jalali, {\it IEEE Photonics Technol. Lett.\/} {\bf 15}, 581
  (2003).

\bibitem{cundiff2010optical}
S.~T. Cundiff, A.~M. Weiner, {\it Nat. Photonics\/} {\bf 4}, 760 (2010).

\bibitem{ye2011photonic}
J.~Ye, {\it et~al.\/}, {\it Opt. Lett.\/} {\bf 36}, 1458 (2011).

\bibitem{cao2021thermo}
Y.~Cao, {\it et~al.\/}, {\it Photonics Res.\/} {\bf 9}, 596 (2021).

\bibitem{schnebelin2019programmable}
C.~Schn{\'e}belin, J.~Aza{\~n}a, H.~Guillet~de Chatellus, {\it Nat. Commun.\/}
  {\bf 10}, 4654 (2019).

\bibitem{khan2010ultrabroad}
M.~H. Khan, {\it et~al.\/}, {\it Nat. Photonics\/} {\bf 4}, 117 (2010).

\bibitem{yao2010arbitrary}
J.~Yao, {\it Nat. Photonics\/} {\bf 4}, 79 (2010).

\bibitem{zhang2020photonic}
W.~Zhang, J.~Yao, {\it Nat. Commun.\/} {\bf 11}, 406 (2020).

\bibitem{ghelfi2014fully}
P.~Ghelfi, {\it et~al.\/}, {\it Nature\/} {\bf 507}, 341 (2014).

\bibitem{marpaung2013integrated}
D.~Marpaung, {\it et~al.\/}, {\it Laser Photonics Rev.\/} {\bf 7}, 506 (2013).

\bibitem{fandino2017monolithic}
J.~S. Fandi{\~n}o, P.~Mu{\~n}oz, D.~Dom{\'e}nech, J.~Capmany, {\it Nat.
  Photonics\/} {\bf 11}, 124 (2017).

\bibitem{marpaung2019integrated}
D.~Marpaung, J.~Yao, J.~Capmany, {\it Nat. Photonics\/} {\bf 13}, 80 (2019).

\bibitem{liu2020integrated}
Y.~Liu, A.~Choudhary, D.~Marpaung, B.~J. Eggleton, {\it Adv. Opt. Photonics\/}
  {\bf 12}, 485 (2020).

\bibitem{fan2008sensitive}
X.~Fan, {\it et~al.\/}, {\it Anal. Chim. Acta\/} {\bf 620}, 8 (2008).

\bibitem{albert2013tilted}
J.~Albert, L.-Y. Shao, C.~Caucheteur, {\it Laser Photonics Rev.\/} {\bf 7}, 83
  (2013).

\bibitem{tan2013microfiber}
Y.~Tan, L.-P. Sun, L.~Jin, J.~Li, B.-O. Guan, {\it Opt. Express\/} {\bf 21},
  154 (2013).

\bibitem{klimov2015chip}
N.~N. Klimov, S.~Mittal, M.~Berger, Z.~Ahmed, {\it Opt. Lett.\/} {\bf 40}, 3934
  (2015).

\bibitem{monzon2011compact}
D.~Monzon-Hernandez, A.~Martinez-Rios, I.~Torres-Gomez, G.~Salceda-Delgado,
  {\it Opt. Lett.\/} {\bf 36}, 4380 (2011).

\bibitem{li2021chip}
A.~Li, Y.~Fainman, {\it Nat. Commun.\/} {\bf 12}, 2704 (2021).

\bibitem{schreiber2023variations}
K.~U. Schreiber, J.~Kodet, U.~Hugentobler, T.~Kl{\"u}gel, J.-P.~R. Wells, {\it
  Nat. Photonics\/} {\bf 17}, 1054 (2023).

\bibitem{newton1993new}
I.~Newton, {\it Am. J. Phys.\/} {\bf 61}, 108 (1993).

\bibitem{choi2008miniature}
H.~Y. Choi, {\it et~al.\/}, {\it Opt. Lett.\/} {\bf 33}, 2455 (2008).

\bibitem{zhang2018fully}
W.~Zhang, J.~Yao, {\it Nat. Commun.\/} {\bf 9}, 1396 (2018).

\bibitem{kim2003optical}
S.-J. Kim, T.-J. Eom, B.~H. Lee, C.-S. Park, {\it Opt. Express\/} {\bf 11},
  3034 (2003).

\bibitem{macleod2010thin}
H.~A. Macleod, H.~A. Macleod, {\it Thin-film optical filters\/} (CRC Press,
  2010).

\bibitem{yang2022wavelength}
C.~Yang, {\it et~al.\/}, {\it Advanced Photonics Res.\/} {\bf 3}, 2100338
  (2022).

\bibitem{wan2021switchable}
C.~Wan, {\it et~al.\/}, {\it Nano Lett.\/} {\bf 22}, 6 (2021).

\bibitem{Cheben:07}
P.~Cheben, {\it et~al.\/}, {\it Opt. Express\/} {\bf 15}, 2299 (2007).

\bibitem{gu1998wavelength}
X.~Gu, {\it Opt. Lett.\/} {\bf 23}, 509 (1998).

\bibitem{geng2013line}
Y.~Geng, X.~Li, X.~Tan, Y.~Deng, Y.~Yu, {\it Opt. Express\/} {\bf 21}, 17352
  (2013).

\bibitem{mohammed2006all}
W.~S. Mohammed, P.~W. Smith, X.~Gu, {\it Opt. Lett.\/} {\bf 31}, 2547 (2006).

\bibitem{Antonio-Lopez:10}
J.~E. Antonio-Lopez, A.~Castillo-Guzman, D.~A. May-Arrioja, R.~Selvas-Aguilar,
  P.~LiKamWa, {\it Opt. Lett.\/} {\bf 35}, 324 (2010).

\bibitem{wu2018advanced}
J.~Wu, T.~Moein, X.~Xu, D.~J. Moss, {\it APL Photonics\/} {\bf 3} (2018).

\bibitem{cohen2018response}
R.~A. Cohen, O.~Amrani, S.~Ruschin, {\it Nat. Photonics\/} {\bf 12}, 706
  (2018).

\bibitem{Vincent_Science21}
V.~Ginis, {\it et~al.\/}, {\it Science\/} {\bf 369}, 436 (2020).

\bibitem{huang1994coupled}
W.-P. Huang, {\it J. Opt. Soc. Am. A\/} {\bf 11}, 963 (1994).

\bibitem{lee2012interferometric}
B.~H. Lee, {\it et~al.\/}, {\it Sensors\/} {\bf 12}, 2467 (2012).

\bibitem{needham2024label}
L.-M. Needham, {\it et~al.\/}, {\it Nature\/} {\bf 629}, 1 (2024).

\bibitem{mohanty2017quantum}
A.~Mohanty, {\it et~al.\/}, {\it Nat. Commun.\/} {\bf 8}, 1 (2017).

\bibitem{weihs1996all}
G.~Weihs, M.~Reck, H.~Weinfurter, A.~Zeilinger, {\it Opt. Lett.\/} {\bf 21},
  302 (1996).

\bibitem{crespi2013integrated}
A.~Crespi, {\it et~al.\/}, {\it Nat. Photonics\/} {\bf 7}, 545 (2013).

\bibitem{seron2023boson}
B.~Seron, L.~Novo, N.~J. Cerf, {\it Nat. Photonics\/} {\bf 17}, 702 (2023).

\bibitem{zhou2022photonic}
H.~Zhou, {\it et~al.\/}, {\it Light Sci. Appl.\/} {\bf 11}, 30 (2022).

\bibitem{zhu2022space}
H.~Zhu, {\it et~al.\/}, {\it Nat. Commun.\/} {\bf 13}, 1044 (2022).

\bibitem{meng2023compact}
X.~Meng, {\it et~al.\/}, {\it Nat. Commun.\/} {\bf 14}, 3000 (2023).

\bibitem{du2024ultracompact}
Z.~Du, {\it et~al.\/}, {\it Sci. Adv.\/} {\bf 10}, eadm7569 (2024).

\end{thebibliography}
\bibliographystyle{Science}

\section*{Methods}
\subsection*{Sample fabrications}
We use the SOI platform (thickness of the silicon device layer: 220 nm) to fabricate the cascaded-mode interferometers. The fabrication processes are as follows: First, the ZEP520A e-beam resist with a thickness of 450 nm is spin-coated on the SOI substrate. Second, we use electron-beam lithography to write the designed structures and immerse after-exposure samples into O-xylene to develop the e-beam resist. Third, reactive-ion etching is used to etch the silicon device layer with a full etching depth of 220 nm, and Remover PG is used to remove all remaining resists. After that, a silicon oxide layer with a 700 nm thickness is deposited on the top of the devices as a protection layer using chemical vapor deposition.

\subsection*{Numerical simulations}
We use the Finite-Difference Time-Domain (FDTD) method (Ansys/Lumerical) to simulate and design our devices, including the grating couplers, parallel waveguide couplers, mode-converting gratings, and cascaded-mode interferometers. In simulations, the size parameters of structures are used the same as in experiments. The refractive index of silicon and silicon oxide is 3.46 and 1.46, respectively.

\subsection*{Optical measurements}
The experiment measurement setup is shown in fig. S1. The laser source is a tunable Santec TSL-550 laser (tunable range: 1500 to 1630 nm, linewidth: 200 kHz, wavelength accuracy: $\pm$ 3 pm). A fiber polarization controller is used to adjust the polarization of the light to reach the maximum coupling efficiency for the fiber-grating coupler, which is designed for transverse electric (TE) polarization. The output power is measured with an optical power meter Santec MPM-212 (power range: $-$80 to 10 dBm, wavelength range: 1250 to 1650 nm).

\section*{Acknowledgments}
We acknowledge support from AFOSR grants FA550-19-1-0352. This work was performed, in part, at the Center for Nanoscale Systems (CNS), a member of the National Nanotechnology Coordinated Infrastructure (NNCI), which is supported by the NSF under award no. ECCS-2025158. CNS is a part of Harvard University. I.-C.B.-C. acknowledges support from PRIMA Grant number 201547 from the Swiss National Science Foundation. V.G. acknowledges support from Research Foundation Flanders under grant numbers G032822N and G0K9322N. M.O. acknowledges funding by the European Union (grant agreement 101076933 EUVORAM). The views and opinions expressed are, however, those of the author(s) only and do not necessarily reflect those of the European Union or the European Research Council Executive Agency. Neither the European Union nor the granting authority can be held responsible for them.

 \section*{Author contributions} 
J.L. initiated the project and conceived the concept of cascaded-mode interferometers. J.L. developed the theory with inputs from I.-C.B.-C., V.G., M.O., and F.C.; J.L., I.-C.B.-C., V.G., and M.O. designed the experiment; J.L. fabricated the devices, carried out the measurements, and analyzed the experimental data; J.L. carried out the numerical simulations. F.C. supervised the project. All authors contributed to the analysis, discussion, and writing of the manuscript.

\section*{Conflict of interest}
The authors declare no other competing interests.

\section*{Data availability}
All data needed to evaluate the conclusions in the paper are present in the main text or the supplementary materials. Additional data are available from the corresponding authors upon reasonable request.

\end{document}